\documentclass[apj]{emulateapj}
\usepackage{adjustbox}
\usepackage{graphicx}
\usepackage{epstopdf}
\usepackage{natbib}
\usepackage[stable]{footmisc}
\usepackage{afterpage}
\usepackage{threeparttablex}
\usepackage{floatrow}
\begin{document}

\slugcomment{Accepted by ApJ}

%opening
\title{Precision ephemerides for gravitational wave searches: II. Cyg X-2}
%\author{Sammanani S. Premachandra$^1$, Duncan K. Galloway$^1$, Danny T. Steeghs$^2$, %Tom Marsh$^2$ and Jorge Casares$^3$ }

\author{Sammanani S. Premachandra\altaffilmark{1}
\ Duncan K. Galloway\altaffilmark{1}  }
\affil{ $Monash$ $Centre$ $for$ $Astrophysics$ $(MoCA)$,
$Monash$ $University$, $VIC$ $3800$}

\author {Jorge Casares\altaffilmark{2,3}
\  Danny T. Steeghs\altaffilmark{4}
\  Thomas R. Marsh\altaffilmark{4}}
%\affil{ $Monash$ $Centre$ $for$ $Astrophysics$ $(MoCA)$,
%$Monash$ $University$, $VIC$ $3800$, $Australia^{1}$$^,$$^{2}$} 

\email{Sammanani.Premachandra@monash.edu}
%\email{\myemail}

\altaffiltext{1}{also School of Physics \& Astronomy, Monash University, VIC 3800, Australia}
%\altaffiltext{2}{also School of Mathematical Sciences, Monash University, VIC 3800, Australia}
\altaffiltext{2}{Instituto de Astrof\'{i}sica de canarias, E-38205, La Laguna, Tenerife, Spain}
\altaffiltext{3}{Departamento de Astrof\'{i}sica, Universidad de La Laguna, E-38205, La Laguna, Tenerife, Spain}
\altaffiltext{4}{Department of Physics, University of Warwick, Coventry, CV4 7AL, UK}

%\affil{1. Monash Center for Astrophysics, Monash University, Clayton, Victoria, 3800, Australia}
%\affil{2. Department of Physics, University of Warwick, Coventry CV4 7AL, UK}
%\affil{3. Department of Physics, IAC, Spain}

\begin{abstract}
\noindent Accreting neutron stars in low-mass X-ray binaries (LMXBs) are candidate high-frequency persistent gravitational wave sources. These may be detectable with next generation interferometers such as Advanced LIGO/VIRGO within this decade. However, the search sensitivity is expected to be limited principally by the uncertainty in the binary system parameters. We combine new optical spectroscopy of Cyg X-2 obtained with the Liverpool Telescope (LT) with available historical radial velocity data, which gives us improved orbital parameter uncertainties based on a 44-year baseline. We obtained an improvement of a factor of 2.6 in the orbital period precision and a factor of 2 in the epoch of inferior conjunction $T_{0}$. The updated orbital parameters imply a mass function of 0.65 $\pm$ 0.01 $M_{\odot}$, leading to a primary mass ($M_{1}$) of 1.67 $\pm$ 0.22 $M_{\odot}$ (for $i=62.5 \pm 4^{\circ}$). In addition, we estimate the likely orbital parameter precision through to the expected Advanced LIGO and VIRGO detector observing period and quantify the corresponding improvement in sensitivity via the required number of templates.

\end{abstract}


\section{Introduction}
\noindent Long intervals ($\sim$ $10^{9}$ year) of accretion onto neutron stars in low-mass X-ray binaries (LMXBs) cause the neutron stars to reach spin rates of many hundreds of times a second \citep{Chakrabarty2003}. Slight geometric distortions ($\sim 10^{-6})$ on the star may lead to a quadrupole mass moment, for example, by a spin-misaligned temperature gradient arising from the deep crust of the neutron star \citep{Bildsten1998}. Any such quadrupole moment will lead to persistent gravitational wave emission at twice the neutron star spin frequency ($\nu_{s}$). 
{If one assumes} that the accretion torque is balanced by the gravitational wave (GW) torque, the expected gravitational wave signal strength at Earth is, 
\small \begin{equation} \label{eq:hsecond}
h_c\approx 4 \times 10^{-27}{R_6^{3/4}\over M_{1.4}^{1/4}}
\left(F\over 10^{-8} \ {\rm erg \ cm^{-2} \ s^{-1}}\right)^{1/2}
\left(300 \ {\rm Hz}\over \nu_s\right)^{1/2},  
\end{equation} 
\normalsize
where $F$ is the observed X-ray flux,  \begin{math} \nu_{s} \end{math} is the spin frequency, $R$ and $M$ are the neutron-star radius and the mass, respectively \citep{Bildsten1998}. Here the strain $h_{c}$ is the fractional change in the length of an interferometer arm (for example). {However, note that there are alternative explanations for the observed spin period distributions (see \citet{Haskell2015} and references therein). }
 
The strongest gravitational wave sources are likely those with short spin frequencies and high mass accretion rates \citep{Bildsten1998}. Unfortunately, spin periods are still unknown for most of the brightest neutron-star binaries. Spin measurements for about 20$\%$ of the known LMXBs have been made by detecting persistent or transient X-ray intensity pulsations {{\citep[e.g.,][]{Watts2012,Patruno2012}}}. However, these phenomena have not been detected in the brightest LMXB sources.  For sources that exhibit twin kHz QPOs, spin frequencies have been estimated from the relationship between the twin kHz QPO separation and the spin frequency; i.e., the kHz QPO separation can be either the spin frequency or half the spin frequency {{\citep{Wijnands2003,Linares2005}}}. However, this may not be a reliable method of estimating the spin frequencies because the separations vary for some of the kHz QPOs \citep{Van2006}. {{Furthermore, comparison studies of the twin kHz QPO separation and spin/burst oscillation frequency showed that there is no direct relationship between both phenomena (\citealt{Yin2007,Mendez2007,Watts2008}; hereafter W08).} Inconsistencies of the relationship across sources has led to conclusions that the frequency difference of the twin kHz QPOs in neutron-star LMXBs cannot be directly linked to the spin frequency \citep{Mendez2007}.}  

Major obstacles in detecting GWs from neutron-star binaries are the large parameter uncertainties of the system parameters (W08) and the neutron-star spin uncertainty of the LMXB system. The parameter space that needs to be searched over is directly proportional to these uncertainties, and demands a large number of possible template models. Therefore, minimizing the parameter space volume through improvements in the parameter uncertainties will contribute to a more sensitive and also a computationally less expensive search.

GW searches have been already made with data from the intitial LIGO detectors \citep{Abb07}, focussing on the brightest of the known LMXBs, Sco X-1. The first search used a fully coherent search from the 2 month second science run (S2) employing 6 hr of data. The second search used a method of cross-correlating the two outputs of the two LIGO detectors utilizing 15 days of data from the fourth science (S4) run \citep{Abbott2007}. These searches were unsuccesful in detecting the signal likely due to the large uncertainties of the model parameters and limited computational power. Another possible cause for the non-detection is that Sco X-1 produces a very low GW amplitude. In our previous paper, we presented our results for Sco X-1, in order to improve the precision of the binary parameters \citep{Galloway2014}.  

After Sco X-1, Cyg X-2 is considered to be the best target for GW searches since it is X-ray bright and has an accessible optical counterpart. This low-mass X-ray binary, discovered in 1965 \citep{Bowyer1965} contains a neutron star and is also known as a Z source \citep*[i.e, the source follows a ``Z" shaped spectral pattern in the X-ray color-color diagram;][]{Hasinger1989}. The neutron star accretes persistently near its Eddington limit, and is a known thermonuclear burst source \citep{Smale1998} which confirms the neutron star nature of the compact object. \citet{Galloway2008} reported a long-term average flux of 11 $\times$ $10^{-9}$ \begin{math} {\rm erg\, cm^{-2}s^{-1}} \end{math} in the 2.5--25 keV band, corresponding to an accretion rate of $>0.8\ \dot{M}_{\rm Edd}$. The estimated distance is given as 11 $\pm$ 2 or 14 $\pm$ 3 kpc, depending on the thermonuclear burst fuel composition \citep{Galloway2008}. {{However, optical observations demonstrated a distance of 7.2 $\pm$ 1.1 kpc \citep{Orosz1999}}. The discrepancy of the measurement with the photospheric radius expansion (PRE) burst distances was earlier reported by \citet{Orosz1999}.} \citet{Wijnands1998} discovered twin kHz QPOs with the $Rossi$ $X$-$ray$ $Timing$ $Explorer$ $(RXTE)$ when the source was in the ``horizontal branch'' (HB) near frequencies of 500 and 860 Hz.
 %They traced this source in the X-ray color-color diagram when the QPOs were present in theHB 
Those authors suggested that the neutron star spin frequency is 346 $\pm$ 29 Hz, by considering the similarity between the QPO peak separation and the neutron star spin frequency. \citet*{Kuulkers1995} placed an upper limit of $10\%$ on burst oscillation amplitudes between 1 and 256 Hz using EXOSAT data, while \citet{Smale1998} placed an upper limit of $2\%$ in the 200-600 Hz frequency range using $RXTE$. {\citet[hereafter C79]{Cowley1979} studied the optical counterpart and derived the first orbital solution}. A refined orbital ephemeris was given by \citep[hereafter C98]{Casares1998} using high-resolution spectroscopy. C98 found values of the binary period $P_{orb}$ = 9.8444 $\pm$ 0.0003 days, the projected semi-amplitude of the donor star orbit, $K_{2}$ = 88.0 $\pm$ 1.4 km s$^{-1}$ and the systemic velocity $\gamma$ = -209.6 $\pm$ 0.8 km s$^{-1}$. The value of $K_{2}$ was challenged by \citet[hereafter E09]{Elebert2009} and in contrast with E09, \citet[hereafter C10]{Casares2010} found a good agreement with the result reported in C98. Assuming $i=62.5 \pm 4^{\circ}$ based on ellipsoidal model fits to B \& V light curves \citep{Orosz1999}, the implied mass for the primary (neutron) star is $M_{1}$ = $1.71$ $\pm$ $0.21$ $M_{\odot}$ (C10). 

The current orbital parameters for Cyg X-2 are based on 13 years of data (C10). Despite the fact that the precision of these parameters is already relatively high, the current ephemeris information will not be accurate enough to conduct these searches in the Advanced LIGO/VIRGO (aLIGO/AdV) era in order to cover the possible parameter space. Hence we observed the optical counterpart of Cyg X-2 in 2010 August--September and 2011 May--August with the Liverpool Telescope and combined these measurements with historical radial velocity (RV) measurements. In this paper we present improved orbital parameters of this binary system based on the (now) 44-year baseline. In addition we will present the estimates of the likely precision of the parameters that can be achieved throughout the aLIGO/AdV period and quantify the improvement in search sensitivity via the number of templates (models).

\medskip
\medskip

\section{Observations and Data Reduction}

\noindent In this section we describe the data gathering and reduction processes for each epoch of data. Here, we compile all the available data and re-analyze several data sets in order to construct a maximal set of radial velocities (RVs). A log of the observations for this paper is presented in Table 1. These observations consist of five epochs of optical spectra, acquired between 1967 and 2011.

We observed Cyg X-2 between 2010 August 2 and 2011 August 1,  with the Fibre-fed RObotic Dual-beam Optical Spectrograph \citep[FRODOspec;][]{Morales2004} on the robotic 2.0 m Liverpool Telescope (LT) at the Observatorio del Roque de Los Muchachos. The spectrograph is fed by a fibre bundle to form an array of 2 $\times$ 12 lenslets each with a field view of 0.83\arcsec on the sky. The high-resolution mode was used to operate the spectrograph, providing a mean dipersion of 0.35 \AA{}/pixel and spectral resolving power of R $\sim$ 5500 in the blue arm. A total of 20 1700s spectra were obtained in 2010 and 43 1700s spectra in 2011, covering the spectral range 3900--5215 \AA{} with a spectral resolution of 55 km s$^{-1}$. In order to carry out the radial velocity analysis, we observed the stellar template HR114 (A7 III, \citealt{Wilson1953}; C10) using the same spectral configuration. The spectral type of Cyg X-2's companion is given as A9 $\pm$ 2 (C98). We de-biased and flat-fielded all the raw images and used the extraction routines from the standard LT frodospec pipeline \citep{Morales2004}.

For the second epoch of data, we used the RV measurements reported by C79, obtained between 1967 July 2 and 1978 August 9 with the 1 m telescope at the Lick Observatory, 2.1 m, the 4 m telescope at Kitt Peak National Observatory (KPNO), and the 1.8 m telescope at Dominion Astrophysical Observatory (DAO) on glass plates. The authors obtained radial velocities by measuring the plates using the oscilloscope display machine, `ARCTURUS' (C79). They measured the velocities of five different absorption lines (H$\beta$, H$\gamma$+H$\delta$, Ca K, ``Metals" and He II) on the plates.

We reanalysed the spectra reported by C98 for a third epoch, obtained using the 4.2m William Herschel Telescope (WHT), equipped with the Intermediate dispersion Spectrograph \& imaging system (ISIS) triple spectrograph (\citealt{Clegg1992} ; C98). A 0.8\arcsec-1.3\arcsec \, slit width resulted in a spectral resolution of 25 km s$^{-1}$. For more details of these observations, refer to C98.

For a fourth epoch, we reanalysed the spectra reported by C10, obtained on the nights of 1999 July 25--26 using the Utrecht Echelle Spectrograph (UES) on the WHT at the Observatorio del Roque de los Muchachos. Ten 1800 s exposures were obtained with the E31 echelle grating and 2 K $\times$ 2 K SITe1 detector, covering the wavelength range 5300--9000 \AA{}. A 1\arcsec \, slit was used, giving spectral resolution of 10 km s$^{-1}$.

For a fifth epoch, we reanalysed 11 spectra reported by C10, taken on the nights of 1999 July 31 and July 2000 with ISIS on WHT. Here, the 1200B grating was adopted on the blue arm covering the wavelength range 3550--6665 \AA{}. We used a 1\arcsec \, slit, resulting in a resolution of 35 km s$^{-1}$. For further details of these observations, see C10.

\begin{table}[tbp]
\begin{center}
\caption{Observing log of Cyg X-2}
\tiny
\begin{tabular}{*{5}{l}}
\hline\noalign{\smallskip}
\hline\noalign{\smallskip}
Date  &  Telescope/    & Number & Exposure & Ref \\
      & Instrument  & of         & Time     \\
      &                 & spectra & (s)         \\
\hline
1967 July 02 -- 1978 Oct 25 & KPNO,DAO,Lick & 59 & -- & [1]\\
1993 Dec 16 -- 1997 Aug 7 & WHT/ISIS & 42 & 1800 & [2] \\
1999 July 25 -- 26 & WHT/UES & 10 & 1800 & [3] \\
1999 July 31/2000 July 9 & WHT/ISIS & 11 & 1800 & [3] \\
2010 Aug 2 -- September 5 & LT/FRODOspec & 20 & 1700 & [4]\\
2011 May 29 -- 2011 Aug 1 & LT/FRODOspec & 43 & 1700 & [4]\\

\hline
\end{tabular}
\begin{tablenotes}
\footnotesize
\tiny
{\bf References.} - [1]. C79, [2]. C98, [3]. C10, [4]. this paper
\end{tablenotes}
\end{center}
\end{table}

\section{Analysis and Results}

\noindent Some of the LT spectra in 2011 showed low signal-to-noise (S/N $\textless$ 10 per pixel) due to the poor weather conditions that were present at the time of the observations. Therefore, we excluded 5 (out of 43 total) lowest quality spectra from our analysis in order to achieve optimal fit parameters. 

The spectra from WHT and LT were normalized by subtracting a low-order spline fit to the continuum, after masking out the main emission and broad Balmer absorption features (Figure 1, top panel). Masking is performed in order to eliminate spectral features not related to the secondary star; in particular, Balmer absorption lines are likely contaminated by ``filled in" emission from the accretion flow. The spectra were rebinned onto a uniform velocity scale of 22 km s$^{-1}$ pixel$^{-1}$. The template star was also rebinned in an identical manner to the Cyg X-2 spectra. Then we broadened the template star HR114 to $V$ $\sin$\ $i$ = 34 km s$^{-1}$ 
to match the width of the secondary photospheric lines (C10).  Finally, individual velocities were extracted through cross correlation with the template star HR114 (Fig. 1).
The cross correlation is calculated by interpolating over masked regions. The lag at which the cross correlation reaches a maximum is identified by a parabolic approximation to three points around the maximum, allowing calculation of the velocity offset and the (statistical) uncertainty. 

We selected only Ca K absorption lines, omitting the contaminated lines (H$\beta$, H$\gamma$+H$\delta$, ``Metals" and He II) from the radial velocity measurements of Cowley's data \citep{Cowley1979}. We calculated the average uncertainty by taking the mean of the absolute residuals of these line measurements.
We plot radial velocity measurements from 1967-1978 observations at KPNO, DAO, Lick, 1993-1997 WHT, 2010 and 2011 Liverpool telescope (LT) in Figure 2. We fitted the RV measurements with a sinusoid via the Levenberg-Marquardt technique implemented in IDL as MPFITFUN, to iteratively search for the best-fit, and obtained a reduced $\chi^2_{\nu}$ value = 4.6 for 174 degrees of freedom (DOF). 
The high $\chi^2$ value likely implies that systematic uncertainties are still present at a significant level. A possible interpretation is that the different sets of data vary in quality, and one or more data sets contribute disproportionately to a poor $\chi^2$ value. To test this hypothesis, we performed sinusoidal fitting routines on each individual data set from each epoch and found that the 2011 measurements from LT contribute disproportionately to the high $\chi^2$ value.
However, the exclusion of LT data do not show any significant effects on the fit parameters, and the best-fit parameters are consistent within the uncertainties at less than 1$\sigma$ level. Therefore, we included our LT measurements from 2011 to obtain the best set of parameters for 
Cyg X-2. In order to obtain a reduced $\chi^2_{\nu}$ value = 1.0 and estimate the conservative parameter uncertainties,  we re-scaled the measurement errors by $\sqrt{4.6}$. Finally, we obtained the following system parameters. \\*

\indent\indent\indent$T_{0}$ = \begin{math}  $2451219.8262\ $  \pm \ $ 0.0087$  \end{math}(HJD)\\
\noindent\indent\indent\indent$P_{orb}$ = \begin{math}  $9.844766\ $  \pm \ $0.000073$ \end{math}  d \\
\noindent\indent\indent\indent$K_{2}$ = \begin{math}  $86.4\ $  \pm \ $0.6$ \ {\rm km\,s^{-1}} \end{math} \\
\noindent\indent\indent\indent$\gamma$ = \begin{math}  $-207.8\ $  \pm \ $0.3$ \ {\rm km\,s^{-1}}  \end{math}\\

\noindent
where $T_{0}$ corresponds to the inferior conjunction of the secondary star, i.e., at $T_{0}$ the companion is closest to Earth. The uncertainties quoted here are 1-$\sigma$ (68\%). Here, the systemic velocity $\gamma$ has been corrected from the radial velocity of HR 114 by adding -10.2 $\pm$0.9 km s$^{-1}$. Simply rescaling the measurement errors by a factor ($\sqrt{4.6}$, this case) suggests that the parameter uncertainties are underestimated. However, there is also a possible contribution of purely systematic errors. Hence, the systematic error contribution should instead be added to the measurement (RV) errors in quadrature. In order to estimate the upper limit contribution of the systematic effect, we added a systematic contribution and varied the magnitude until we reach a reduced $\chi^2$ value = 1.0. We found that the systematic contribution is 7.9 km s$^{-1}$ and with this value, the resulted system parameters within the uncertainties are consistent at 1$\sigma$ compared to the parameters obtained by rescaling the measurement errors.

\begin{figure}%[H]
\centering
\includegraphics[width=8.5cm,height=6cm]{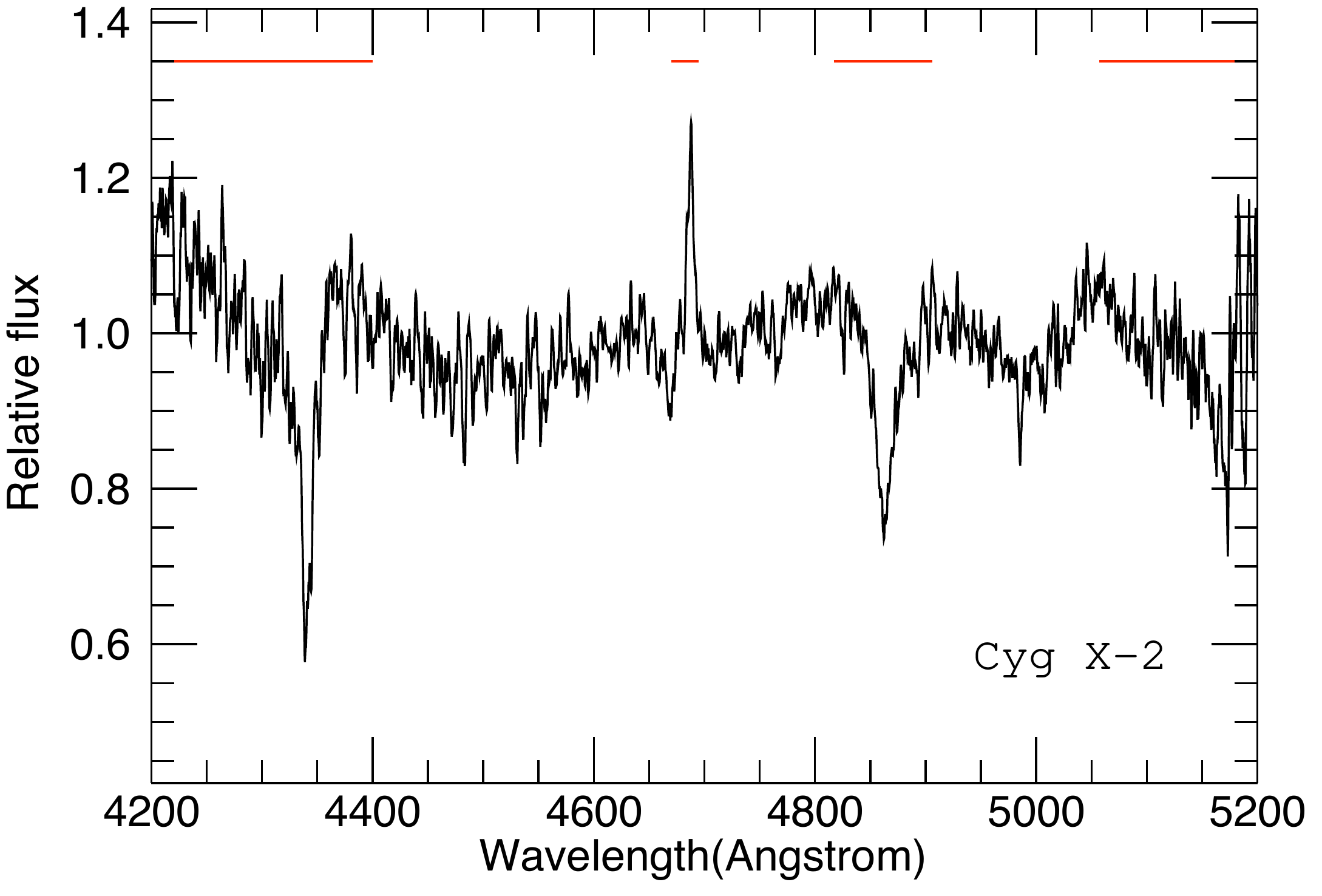}

 \centering
 \includegraphics[width=8.5cm,height=6cm]{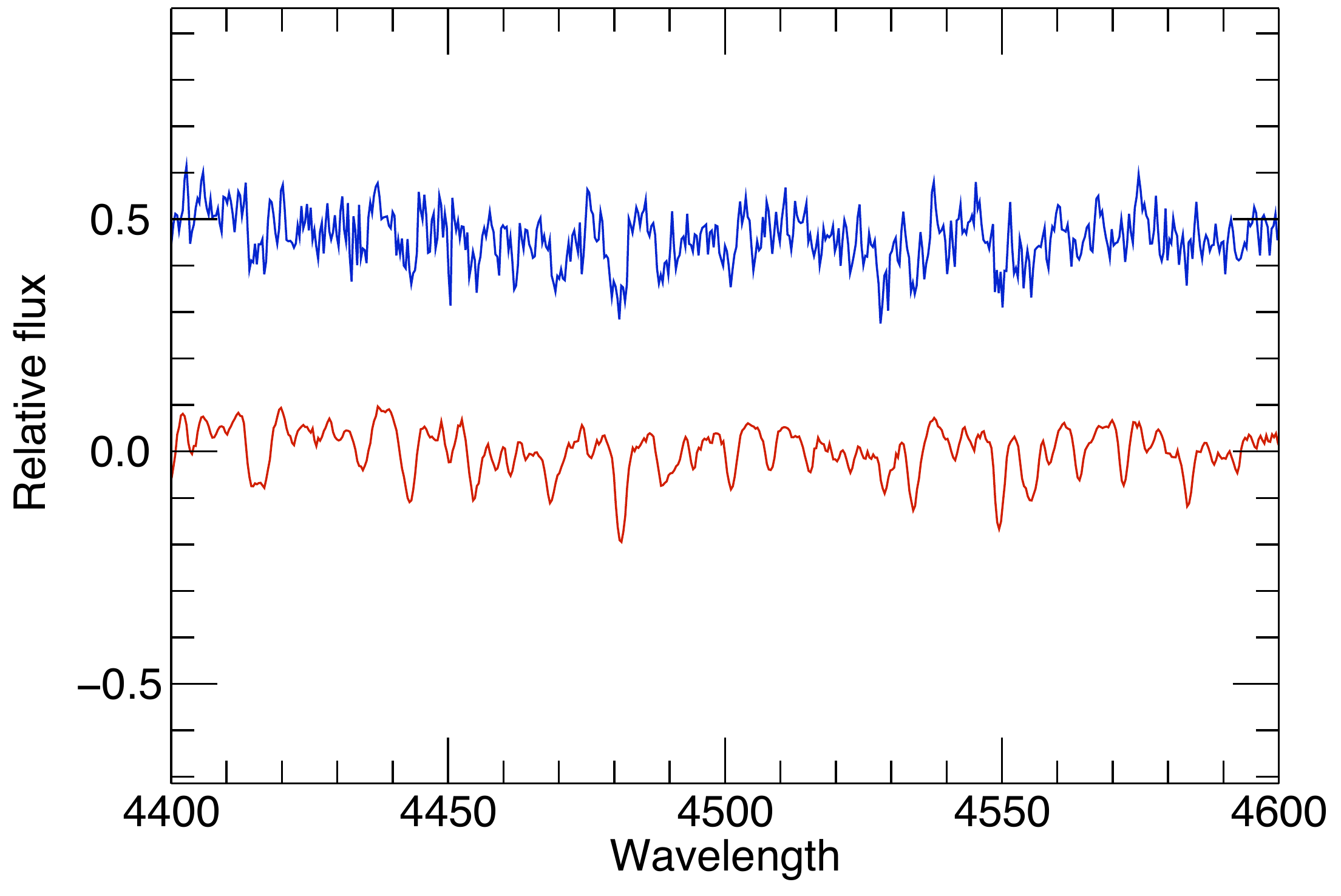}
\caption{%Normalized flux is plotted as a function of wavelength for Cyg X-2 \& HR 114. 
Upper panel showing the averaged spectrum of Cyg X-2 from the 2011 LT data, plotted with horizontal red lines to indicate the wavelength ranges of masked (excluded) regions from the cross-correlation. In the lower panel we show a selected region of the template HR 114 (red), and a section of an averaged Cyg X-2 spectrum in the rest frame of the donor (blue). The Cyg X-2 averaged spectrum is shifted upwards arbitrarily for clarity. }
 \label{Fig 1:}
\end{figure}
%\newpage
%\noindent 

We verified the uncertainties for each parameter using the bootstrap simulation technique of drawing samples randomly from the observed RV measurements, and creating 200 mock data sets. Then the Levenberg-Marquardt algorithm is applied to this data set using the best-fit solution for the real RV measurements as the initial guess. This produced consistent 1$\sigma$ parameter uncertainties. 

The above results show an improvement of factor of 2.6 in the precision of the orbital period error and factor of 2 in the phase zero error, compared to the previous values, 0.00019 (days) and 0.018 (days) of C10. The updated values of the orbital period ($P_{orb}$), velocity amplitude ($K_{2}$) and the systemic velocity ($\gamma$) are consistent (at the 1.5$\sigma$ level) with that of C10. We subtracted 167 orbital cycles from $T_{0}$ to compare with the ephemeris of C10. We obtain fractional days of 0.17 compared to 0.148, which is consistent at the 1.5$\sigma$ level. In addition to this, we find the updated values of the system's mass ratio ($q$) and the neutron mass ($M_{1}$) are consistent to within the uncertainties with the previous value of C10 at 1$\sigma$ (68$\%$) confidence level. The best-fit orbital parameters are listed in Table 2. We varied the initial parameter value of $T_{0}$ to obtain the smallest possible cross-term in the covariance matrix, $V(P_{orb},T_{0})$ giving the cross-term between $P_{orb}$ and $T_{0}$. 

The measured velocity amplitude K, might be a slight overestimate of the velocity amplitude of the companion's center of mass, due to quenching of the absorption features in the heated side of the donor star \citep{Wade1988}. As a result of irradiation, the shape of the radial velocity curve may be distorted from a pure sine wave. In general, this distortion would manifest as an apparent eccentricity in the fits, which should be measurable. Therefore, we also tried to fit an elliptical orbit to the same set of data and found no significant eccentricity with a formal best fit of e = 0.02 $\pm$ 0.03 (with a reduced $\chi^2$ of 4.6), which suggests irradiation is insignificant at this level. Hence, we conclude that a circular orbit yields the best description of the RV measurements and thus we assume the measured K corresponds to the center of mass of the donor star. \citep[e.g.,][]{Davey1992}.

Combined with our new values of $K_{2}$ and $P_{orb}$, we find the mass function value of \\ \begin {equation} f(M)  =  \frac {M_{1} \sin^{3}\ i} {(1+q)^{2}}  = \frac {PK_{2}^{3}}{2 \pi G} =  0.65 \pm 0.01 M_{\odot} \end{equation} Using the value for the rotational broadening of the secondary star, $V$ $\sin$\ $i$ = 33.7 $\pm$ 0.9 ${\rm km\,s^{-1}}$ (C10) and our new value of $K_{2}$, we find a mass ratio of $q$ = 0.34 $\pm$ 0.01. Hence, we calculate $M_{1}$ $\sin^{3}$\ $i$ = 1.17 $\pm$ 0.03 $M_{\odot}$. Assuming $i=62.5 \pm 4^{\circ}$ \citep{Orosz1999}, we obtain a primary mass of $M_{1}$ = 1.67 $\pm$ 0.22 $M_{\odot}$, and a secondary mass of $M_{2}$ = 0.56 $\pm$ 0.07 $M_{\odot}$. We point out that the assumed $V$ $\sin$\ $i$ and inclination do not affect the derived ephemeris in this paper.

\begin{figure}%[H]
 \centering
 \includegraphics[width=8.5cm,height=6cm]{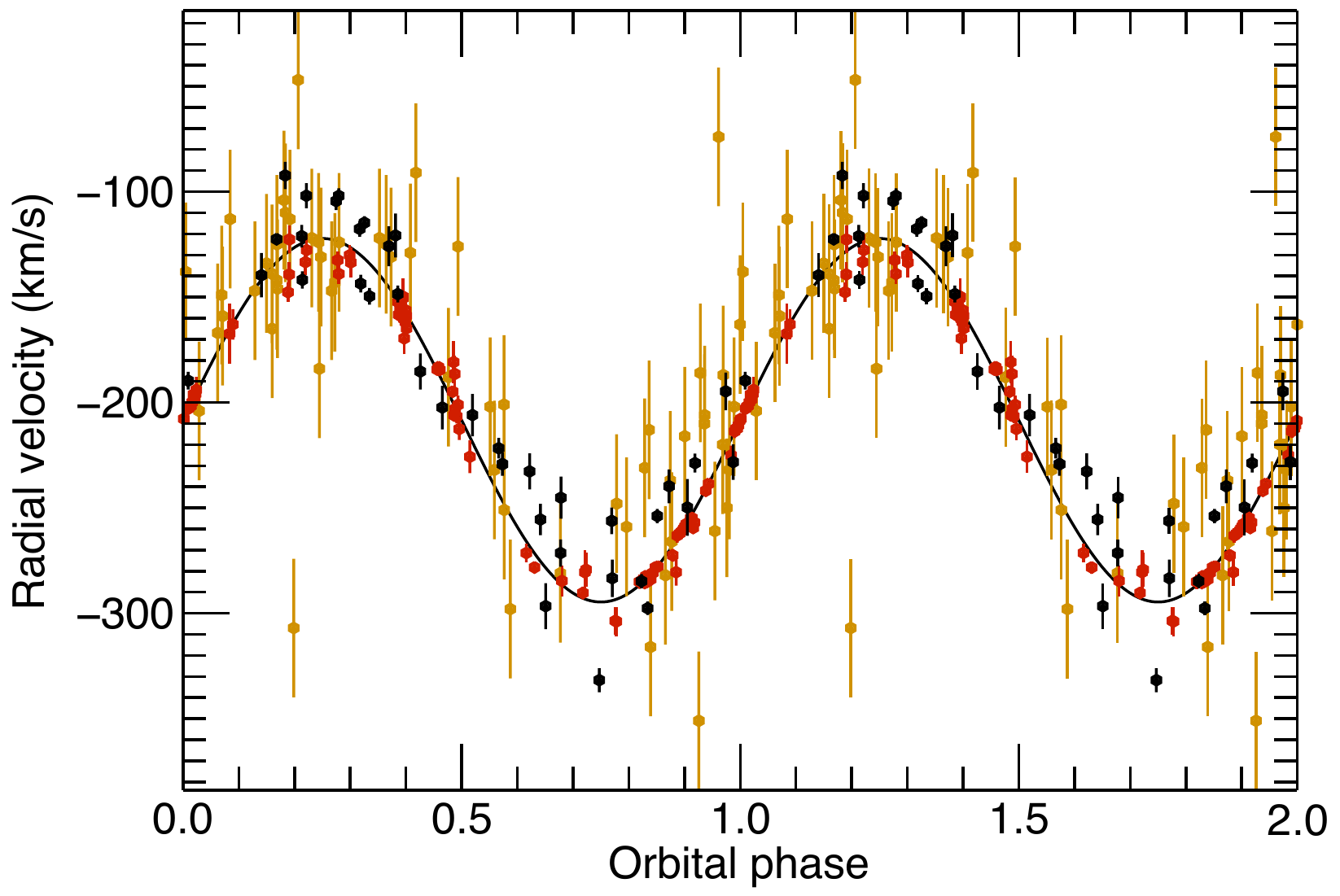}
  %bowen_lines.ps: 612x792 pixel, 72dpi, 21.59x27.94 cm, bb=0 0 612 792
 
\centering
 \includegraphics[width=8.5cm,height=6cm]{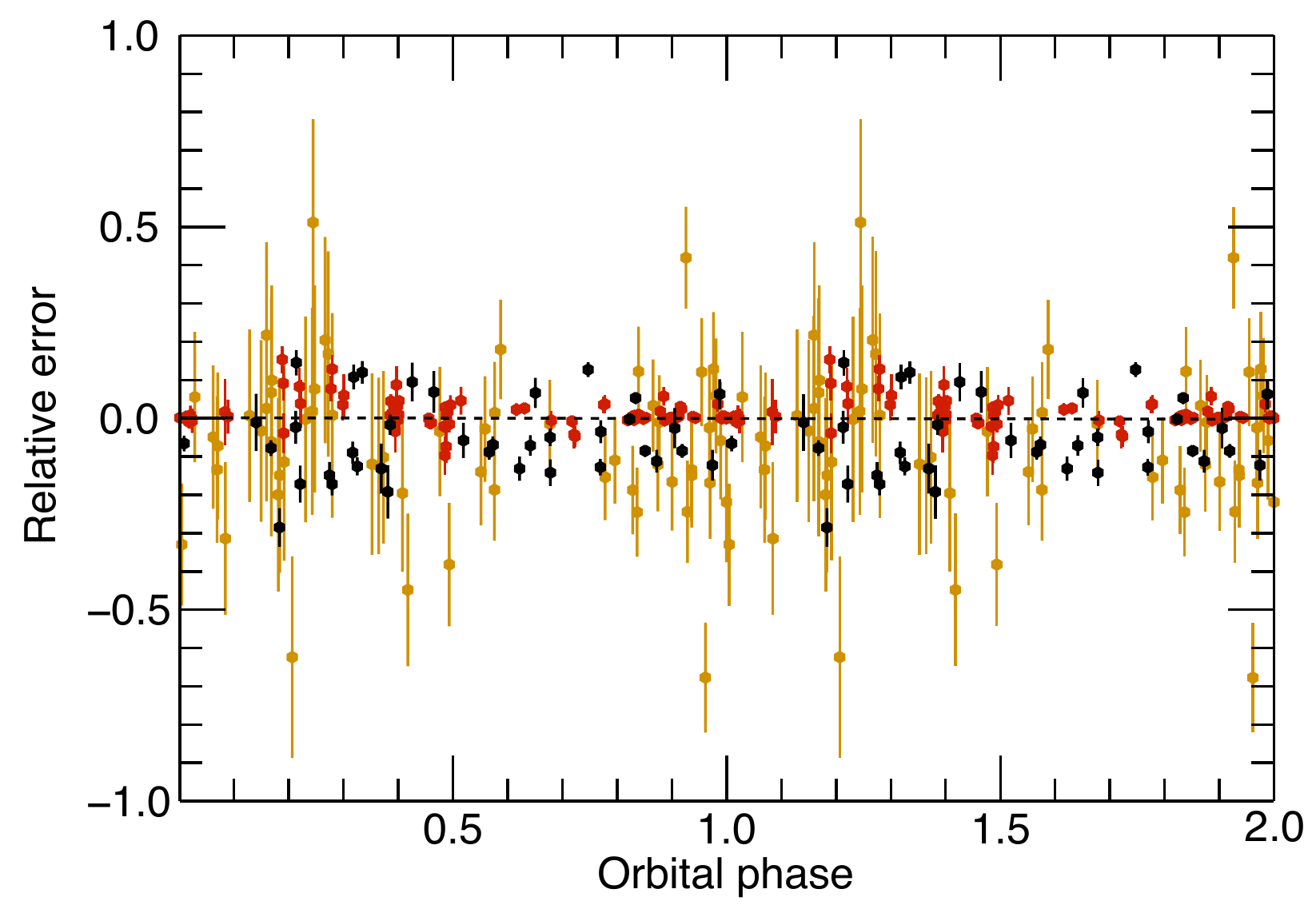}
\caption{Radial velocities for Cyg X-2 as a function of orbital phase from 1976-1978 KPNO, DAO and Lick, 1993-1997 WHT, 2010 \& 2011 LT data. The data are plotted twice for clarity. In the upper panel, we show the radial velocities with the orbital model of \citet{Casares2010} overplotted as a black solid line. Measurements from C79 are shown in yellow, while those from WHT \& LT are shown in red \& black, respectively. The lower panel shows the relative error, i.e., (data - model)/model. Error bars indicate the statistical (1-$\sigma$) uncertainty.}
 \label{Fig 2:}
\end{figure}

%\noindent 
\newpage

\section{Prospects for GWs detectability}
\subsection{Precision estimates throughout Advanced LIGO (aLIGO)/VIRGO (AdV) observations}

\noindent Observing runs for GW searches are planned to commence in 2015, and will run through to 2022+. Estimated run duration for 2015, 2016-17 and 2017-18, are 3, 6 and 9 months, respectively. Starting from 2019, more extended runs are expected take place with aLIGO and AdV per every year. In addition, LIGO India is also expected be included in the observing schedule from 2022 onwards \citep{LIGO2013}.

The inferior conjunction value for the future gravitational wave searches ($T_{n}$) will depend on  the uncertainties on both $T_{0}$ and $P_{orb}$ when aLIGO/VIRGO observations commence. We consider here how the error on $T_{0}$ may be projected in time to any given epoch. Given the epoch for inferior conjunction, $T_{n}$ = $nP_{orb}$ + $T_{0}$, following \citet{Galloway2014}, we derive the projected uncertainty for $T_{n}$ as, 

\begin{equation}  \sigma_{\rm n}^{\rm 2} =  n^{\rm 2} \sigma_{P_{\rm orb}}^{\rm 2} +   \sigma T_{0}^{2} + 2nV(P_{orb},T_{0})\end{equation}

\noindent where, $\sigma_{P_{\rm orb}}$ and $\sigma_{T_{\rm 0}}$ are the uncertainties in the $P_{orb}$ and $T_{0}$, respectively and $V(P_{orb},T_{0})$ is the cross-term of the covariance matrix. Substituting the values from Table 2, we express the error on $T_{n}$ as follows.

\begin{equation} \rm \sigma_{\rm 0},_{\rm t}  \approx [1.3 \times 10^{-8}(t/1\, d)^2 + 3.5 \times 10^{-8}(t/1\,d) + 7.5 \times 10^{-5}]^{1/2} \,d     \end{equation}
\begin{equation} \rm = [1.7 \times 10^{-3}(t/1\, yr)^2 + 1.3 \times 10^{-5}(t/1\,yr) + 7.5 \times 10^{-5}]^{1/2} \,d     \end{equation}

\noindent Here, \begin{math} \rm \sigma_{\rm 0},_{\rm t} \end{math} is the uncertainty in the epoch of inferior conjunction at the time of t in days since $T_{0}$. The evolution of \begin{math} \rm \sigma_{\rm 0},_{\rm t} \end{math} is shown as a function of time in Figure 3. In the absence of additional epochs of observations, $\sigma_{P_{\rm orb}}$ term will grow linearly due to the factor, \\ $4.1$ $\times$ $10^{-2}(t/1\,yr)$\,d. The effective uncertainty will grow approximately as a function of time, 2$\times$ in 2018 and 3$\times$ or greater in/after 2022 compared the current error level.

The above estimated values show that additional epochs of RV measurements will be needed to further refine the system parameters in order to improve the future GW searches. Based on the following assumptions, we also carried out simulations to estimate the effect of additional epochs of RV measurements. First, that the uncertainty on $T_{0}$ decreases as the total number of observations increases. Second, that the uncertainty on $P_{orb}$ decreases as the total span of the observations increases. We generated 6 epochs of 100 simulated radial velocity measurements in total. We considered each observing epoch to consist of 15 random observations of Liverpool Telescope between 2015 and 2025. We have estimated the parameter uncertainties by combining all existing data with the simulated 100 RV measurements. Simulations allow us to improve the uncertainty on the orbital period down to a level of approximately $4.5$ $\times$ $10^{-5}$ d and maintain the uncertainty on $T_{0}$ at or below the current level ($\approx$ 10$^{-3}$ d) throughout the aLIGO/AdV observations.

%\begin{tabular}{*{5}{c}}
%\newpage
%Table 2: Updated orbital parameters for Cyg X-2
\begin{table}[tbp]
\begin{center}
\caption{Orbital parameters for Cyg X-2}
\begin{tabular}{*{3}{l}}
\hline\noalign{\smallskip}
\hline\noalign{\smallskip}
parameter & Value & Units\\
\hline

$\gamma$ & \begin{math}  $-207.8\ $  \pm \ $0.3$   \end{math} & \ ${\rm km\,s^{-1}}$\\
$K_{2}$ & \begin{math}  $86.4\ $  \pm \ $0.6$   \end{math} &  \ ${\rm km\,s^{-1}}$\\
$T_{0}$ & \begin{math}  $2451219.8262\ $  \pm \ $0.0087$   \end{math} & \ HJD\\
	& \begin{math} $602668183\ $ \pm \ 750 \end{math} & \ GPS seconds\\ 
        & 1999 February 10 at 07:49:43  & \ UTC\\
$P_{orb}$ & \begin{math}  $9.844766\ $  \pm \ $0.000073$   \end{math} & \ d\\
$V(P_{orb},T_{0})$ & 1.15 $\times$ $10^{-8}$ & \ d$^{2}$ \\
\hline
\end{tabular}

\end{center}
\end{table}

\medskip
\medskip

%\newpage
\subsection{Template Calculations}
\noindent In general, multiple templates (models) are required to carry out GW searches in parameter space due to the parameter uncertainties. Therefore, calculating the number of required templates is a measure of the computational cost involved. The size of the parameter space is given by the volume measure in a lattice grid (e.g., W08), i.e., the total volume of the parameter space is divided by the volume of each unit cell. The mismatch between the template and the signal is measured by the fractional loss in the S/N of the two waveforms in the space. \citet*{Dhur2001} were the first to carry out a detailed study on the parameter space metric for a coherent matched filter search for a neutron star in a binary orbit, while \citet{Abb07} carried out the first GW search on Sco X-1. Once the best set of system parameters are available, the next important step is to use this best possible parameters to check the effect on number of templates (models) needed for the GW searches. Here, we use the template counting equations derived by W08 to determine the number of templates required at the time of aLIGO/AdV searches. 

%W08 have employed a total mismatch,\\ m = 30$\%$ in their paper and we use the same m %value for comparison.

The number of templates for a joint search of $P_{orb}$,$T_{0}$ space is,

 \begin{equation} N_{\rm P_{orb}},_{\rm T_{0}} \propto \sigma[P_{orb}^{-2}]\sigma[T_{0}] \end{equation}

\medskip

%The number of templates corresponding to the long observation search., i.e, \\$T_{\rm obs}$ %$\gg$ $P_{\rm orb}$ and as a reference value $T_{\rm obs}$ = 2 $yr$ are given by,
\medskip

%\begin{eqnarray}
% N_{P_{\rm orb}} &=& 2 \times \sigma P_{\rm orb} \frac{\sqrt{8}\pi^2 a_{\rm
 %     p}^{\rm max} \times 2 \times\nu_{\rm max} T_{\rm obs} }{ \sqrt{3m}(P_{\rm orb}^{\rm %min})^2}\,,\\
% N_{T_{\rm 0}} &=& 2 \times\sigma T_{\rm 0}\frac{\sqrt{8}\pi^2  a_{\rm
 %     p}^{\rm max} \times 2 \times\nu_{\rm max}}{  P_{\rm orb}^{\rm min}\sqrt{m}} \,.
 %\end{eqnarray}

%$\nu$ is the neutron star spin frequency and $a_{p}$ is the semi major axis of the neutron %star.Here, the first factor of 2 in each equation corresponds to calculating the total range for %the parameter uncertainty while the other factor 2 defines the kHz QPOs frequency separation.   

\noindent where, $\sigma$ is the uncertainty for each system parameter, $P_{orb}$ is the orbital period, $T_{0}$ is the inferior conjunction of the companion \footnote[5]{We point out that the $P_{orb}$ uncertainty value for Cyg X-2 in W08 paper (in Table 4), is incorrect and it should be $\pm$ 7.2$\times$10$^{-3}$~h instead of $\pm$ 8.3$\times$10$^{-8}$ h. Hence, the corresponding corrected number of templates ($N_{P \rm orb}$) in W08 should be 336}. In this paper, we quantify the expected sensitivity improvement based on the fractional reduction in the number of templates that derives from the reduction in the uncertainty in $P_{orb}$ and $T_{0}$. The fractional reduction in the number of templates for the parameters listed in Table 2 is a factor of 5. However, this factor is an underestimate of the improvement in number of templates for future searches because this factor is based only on the parameters derived in this paper. Hence, the effective uncertainty in the orbital parameters must be propagated through to aLIGO observing period including the contribution from the other parameters. Based on our simulated measurements, we expect the improvement in the templates number to be at a factor of 20 or better when aLIGO/AdV observations commence from 2015 onwards.

{{Recent studies have shown a more feasible method of estimating the search sensitivity for LMXBs by employing a mock-data challenge \citep{Messenger2015}.
%{\bf{Recently, \citet{Messenger2015} employed a mock-data challenge for the case of Sco X-1, %which is a more feasible method of estimating the search senistivity for LMXBs. 
We expect to adopt the same method for Cyg X-2 and derive the sensitivity estimates in order to illustrate the improvement more explicitly in a future paper.  }} 

\medskip
\medskip

%\newpage

\begin{figure}[ht]%[H]
 \centering
 \vspace{0.25cm}
 \includegraphics[width=8.5cm,height=6cm]{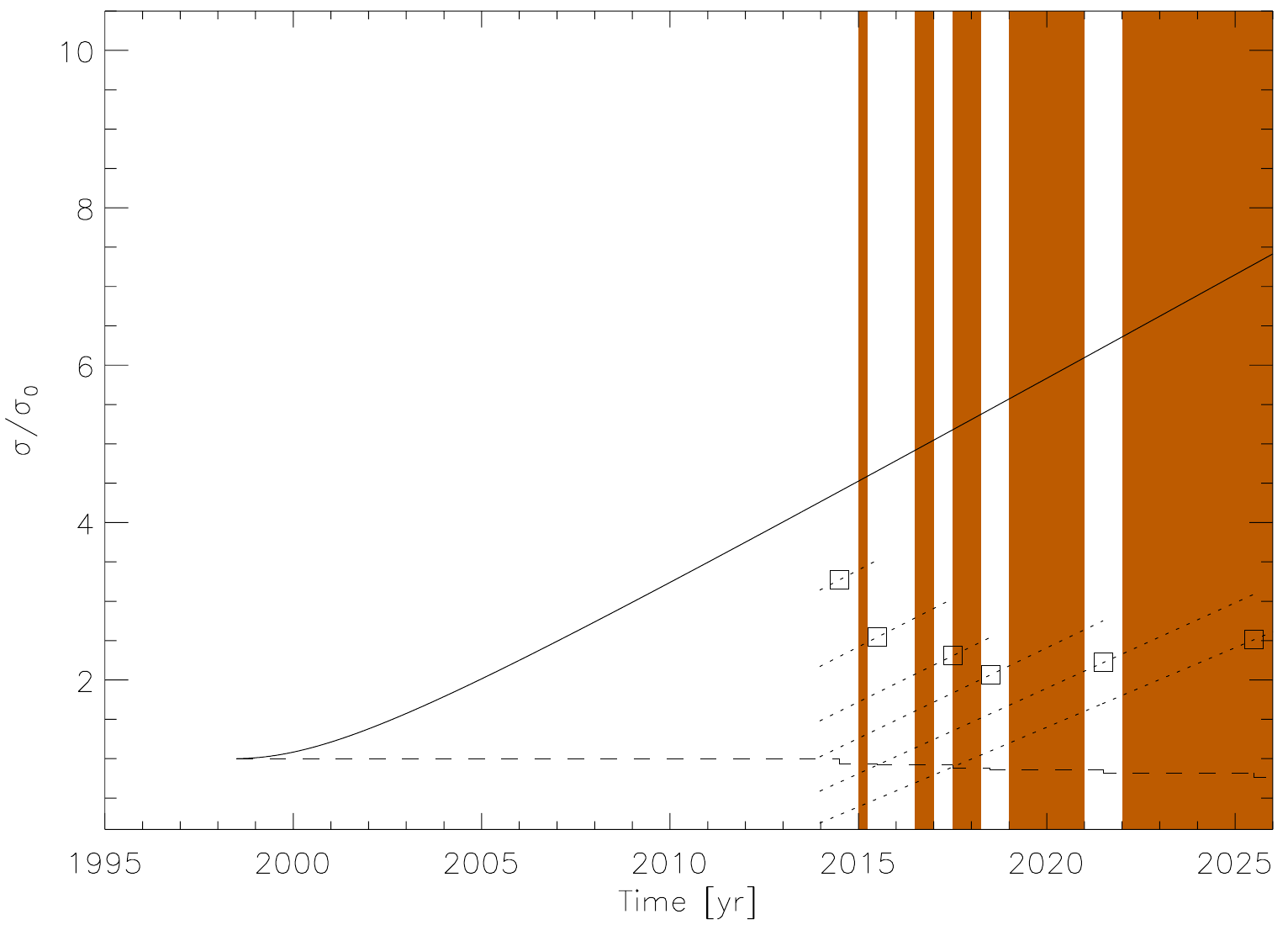}
\caption{The projected uncertainty throughout the aLIGO observing period (shaded regions) is plotted as a function of time. The solid line denotes the approximately linear growth of the effective error throughout the aLIGO observing period (from 2015 onwards) in the absence of the additional epochs of optical data. The open squares describe the approximate effect on the additional observing epochs and the $T_{0}$ uncertainty is represented by the dotted curves. The estimated orbital period uncertainty ($\sigma$$P_{orb}$) is denoted by a dashed line.  }
 \label{Fig :}
\end{figure}

\section{Conclusions}

We have obtained an improved set of orbital parameters based on a 44-year baseline for the neutron-star binary Cyg X-2. The precision of the orbital period is based on the long-baseline of observations and as well as how precisely the orbital phase can be measured at any given epoch. 
Our new set of optical data supports the velocity amplitude of the companion reported in C10 over the value suggested by E09. These new measurements resulted in an improvement in the precision of the orbital period uncertainty ($\sigma_{P_{\rm orb}}$) by a factor of 2.6 and orbital phase uncertainty ($\sigma_{T_{\rm 0}}$) by a factor of 2 compared to those reported by C10. The updated $P_{orb}$ and $K_{2}$ give a mass function of 0.65 $\pm$ 0.01 $M_{\odot}$ (1 $\sigma$) and a primary mass of 1.67 $\pm$ 0.22 $M_{\odot}$. These are also consistent with that of C10. 

We also examined the effect of the newly derived system parameters on future GW searches through to 2025.  We applied the updated orbital parameters to equation (5) and obtained an improvement in the number of templates by a factor of 5 (compared to that of C10). However, these factors are  based on the orbital parameters that are determined in this paper (Table 2). We expect $T_0$ uncertainty along with the other parameter uncertainties to be improved further at the time of LIGO observations commence.  Therefore, we quantified the number of templates for the epoch of future GW searches and achieved an overall improvement of a factor of 20 or better. This improvement demonstrates that more precise system parameter uncertainties can make a substantial difference to the number of templates searched in parameter space. Thus, providing an accurate measurements of the orbital parameters and the absolute phase of the binary will permit us to obtain a step change in sensitivity for GW searches in future. However, we point out that this is a conservative estimate and expect to further improve on the template requirement predictions with the help of additional observing epochs. 
%Therefore, we propose the following observing plan.

%Based on our 2011 observations, we will request longer integration times (e.g., 1700 s) to obtain spectra. These spectra will be taken using PATT and CAT with LT telescope on evenly distributed nights in every year. 

Additional observations (e.g., Fig 3) will enable us to maintain a long baseline of observations, hence refine the orbital parameters for future GW searches and other spectroscopic studies such as constraining the neutron star mass. {{Even though the relationship between the pulsations and orbital variations is completely independent of each other (e.g., W08), the lack of precise knowledge of the spin frequency of this source contributes to a larger number of templates for gravitational wave searches. Hence, more deeper searches for pulsations for this source should also be carried out in order to determine more precise neutron-star spin frequencies \citep[e.g.,][] {Mess2015}. This would provide the most substantial improvement in the search sensitivity. }}   

%Additional PATT observing campaigns also will allow us understand and compare the systematic %effects (low signal-to-noise) that was present in 2011 PATT spectra. 

\begin{acknowledgments}

This project was funded partly by the Monash-Warwick Strategic Funding Initiative. S.P. is supported by Faculty of Science Dean's Postgraduate Research Scholarship. D.K.G. acknowledges the support of an Australian Research Council Future Fellowship (project FT0991598). 
J.C. acknowledges support by the Spanish Ministerio de Econom\'{i}a y Competividad (MINECO) under grant AYA2010--18080. D.S. acknowledges support from STFC through an Advanced Fellowship (PP/D005914/1) as well as grant ST/I001719/1. TRM and D.S. were partially supported under grants from the UK's Science and Technology Facilities Coulcil (STFC), grant number ST/L000733/1. The LT is operated on the island of La Palma by Liverpool John Morres University in the Spanish Observatorio del Roque de los Muchachos of the Instituto de Astrofisica de Canarias with financial support from the UK Science and Technology Facilities Council.

\end{acknowledgments}

%\newpage
\bibliographystyle{apj}

%\bibliography{Abb07,Abbott2007,Messenger2015,Mess2015,Yin2007,Mendez2007,Patruno2012,Bildsten1998,Dib2005,Chakrabarty2003,Galloway2014,Elebert2009,Morales2004,Watts2012,Orosz1999,Watts2008,Wij1998,Wijnands1998,Galloway2002,Galloway2008,Hasinger1989,Kuulkers1995,Casares1998,Casares2010,Bowyer1965,Cowley1979,Smale1998,Clegg1992,Horne1986,Dhur2001,Wijnands2003,Wade1988,Davey1992,Linares2005,Van2006,Wilson1953,LIGO2013,Haskell2015}

\begin{thebibliography}{34}
\expandafter\ifx\csname natexlab\endcsname\relax\def\natexlab#1{#1}\fi

\bibitem[{{Abbott} {et~al.}(2007{\natexlab{a}}){Abbott}, {Abbott}, {Adhikari},
  {Agresti}, {Ajith}, {Allen}, {Amin}, {Anderson}, {Anderson}, {Arain}, \&
  et~al.}]{Abb07}
{Abbott}, B., {et~al.} 2007{\natexlab{a}}, \prd, 76, 082001

\bibitem[{{Abbott} {et~al.}(2007{\natexlab{b}}){Abbott}, {Abbott}, {Adhikari},
  {Agresti}, {Ajith}, {Allen}, {Amin}, {Anderson}, {Anderson}, {Arain}, \&
  et~al.}]{Abbott2007}
---. 2007{\natexlab{b}}, \prd, 76, 082003

\bibitem[{{Bildsten}(1998)}]{Bildsten1998}
{Bildsten}, L. 1998, \apjl, 501, L89

\bibitem[{{Bowyer} {et~al.}(1965){Bowyer}, {Byram}, {Chubb}, \&
  {Friedman}}]{Bowyer1965}
{Bowyer}, S., {Byram}, E.~T., {Chubb}, T.~A., \& {Friedman}, H. 1965, Annales
  d'Astrophysique, 28, 791

\bibitem[{{Casares} {et~al.}(1998){Casares}, {Charles}, \&
  {Kuulkers}}]{Casares1998}
{Casares}, J., {Charles}, P.~A., \& {Kuulkers}, E. 1998, \apjl, 493, L39

\bibitem[{{Casares} {et~al.}(2010){Casares}, {Gonz{\'a}lez Hern{\'a}ndez},
  {Israelian}, \& {Rebolo}}]{Casares2010}
{Casares}, J., {Gonz{\'a}lez Hern{\'a}ndez}, J.~I., {Israelian}, G., \&
  {Rebolo}, R. 2010, \mnras, 401, 2517

\bibitem[{{Chakrabarty} {et~al.}(2003){Chakrabarty}, {Morgan}, {Muno},
  {Galloway}, {Wijnands}, {van der Klis}, \& {Markwardt}}]{Chakrabarty2003}
{Chakrabarty}, D., {Morgan}, E.~H., {Muno}, M.~P., {Galloway}, D.~K.,
  {Wijnands}, R., {van der Klis}, M., \& {Markwardt}, C.~B. 2003, \nat, 424, 42

\bibitem[{{Clegg} {et~al.}(1992){Clegg}, {Cordes}, {Simonetti}, \&
  {Kulkarni}}]{Clegg1992}
{Clegg}, A.~W., {Cordes}, J.~M., {Simonetti}, J.~M., \& {Kulkarni}, S.~R. 1992,
  \apj, 386, 143

\bibitem[{{Cowley} {et~al.}(1979){Cowley}, {Crampton}, \&
  {Hutchings}}]{Cowley1979}
{Cowley}, A.~P., {Crampton}, D., \& {Hutchings}, J.~B. 1979, \apj, 231, 539

\bibitem[{{Davey} \& {Smith}(1992)}]{Davey1992}
{Davey}, S., \& {Smith}, R.~C. 1992, \mnras, 257, 476

\bibitem[{{Dhurandhar} \& {Vecchio}(2001)}]{Dhur2001}
{Dhurandhar}, S.~V., \& {Vecchio}, A. 2001, \prd, 63, 122001

\bibitem[{{Elebert} {et~al.}(2009){Elebert}, {Callanan}, {Torres}, \&
  {Garcia}}]{Elebert2009}
{Elebert}, P., {Callanan}, P.~J., {Torres}, M.~A.~P., \& {Garcia}, M.~R. 2009,
  \mnras, 395, 2029

\bibitem[{{Galloway} {et~al.}(2008){Galloway}, {Muno}, {Hartman}, {Psaltis}, \&
  {Chakrabarty}}]{Galloway2008}
{Galloway}, D.~K., {Muno}, M.~P., {Hartman}, J.~M., {Psaltis}, D., \&
  {Chakrabarty}, D. 2008, \apjs, 179, 360

\bibitem[{{Galloway} {et~al.}(2014){Galloway}, {Premachandra}, {Steeghs},
  {Marsh}, {Casares}, \& {Cornelisse}}]{Galloway2014}
{Galloway}, D.~K., {Premachandra}, S., {Steeghs}, D., {Marsh}, T., {Casares},
  J., \& {Cornelisse}, R. 2014, \apj, 781, 14

\bibitem[{{Hasinger} \& {van der Klis}(1989)}]{Hasinger1989}
{Hasinger}, G., \& {van der Klis}, M. 1989, \aap, 225, 79

\bibitem[{{Haskell} {et~al.}(2015){Haskell}, {Priymak}, {Patruno},
  {Oppenoorth}, {Melatos}, \& {Lasky}}]{Haskell2015}
{Haskell}, B., {Priymak}, M., {Patruno}, A., {Oppenoorth}, M., {Melatos}, A.,
  \& {Lasky}, P.~D. 2015, \mnras, 450, 2393

\bibitem[{{Kuulkers} {et~al.}(1995){Kuulkers}, {van der Klis}, \& {van
  Paradijs}}]{Kuulkers1995}
{Kuulkers}, E., {van der Klis}, M., \& {van Paradijs}, J. 1995, \apj, 450, 748

\bibitem[{{LIGO Scientific Collaboration} {et~al.}(2013){LIGO Scientific
  Collaboration}, {Virgo Collaboration}, {Aasi}, {Abadie}, {Abbott}, {Abbott},
  {Abbott}, {Abernathy}, {Accadia}, {Acernese}, \& et~al.}]{LIGO2013}
{LIGO Scientific Collaboration} {et~al.} 2013, ArXiv:1304.0670

\bibitem[{{Linares} {et~al.}(2005){Linares}, {van der Klis}, {Altamirano}, \&
  {Markwardt}}]{Linares2005}
{Linares}, M., {van der Klis}, M., {Altamirano}, D., \& {Markwardt}, C.~B.
  2005, \apj, 634, 1250

\bibitem[{{M{\'e}ndez} \& {Belloni}(2007)}]{Mendez2007}
{M{\'e}ndez}, M., \& {Belloni}, T. 2007, \mnras, 381, 790

\bibitem[{{Messenger} \& {Patruno}(2015)}]{Mess2015}
{Messenger}, C., \& {Patruno}, A. 2015, \apj, 806, 261

\bibitem[{{Messenger} {et~al.}(2015){Messenger}, {Bulten}, {Crowder},
  {Dergachev}, {Galloway}, {Goetz}, {Jonker}, {Lasky}, {Meadors}, {Melatos},
  {Premachandra}, {Riles}, {Sammut}, {Thrane}, {Whelan}, \&
  {Zhang}}]{Messenger2015}
{Messenger}, C., {et~al.} 2015, ArXiv:1504.05889

\bibitem[{{Morales-Rueda} {et~al.}(2004){Morales-Rueda}, {Carter}, {Steele},
  {Charles}, \& {Worswick}}]{Morales2004}
{Morales-Rueda}, L., {Carter}, D., {Steele}, I.~A., {Charles}, P.~A., \&
  {Worswick}, S. 2004, Astronomische Nachrichten, 325, 215

\bibitem[{{Orosz} \& {Kuulkers}(1999)}]{Orosz1999}
{Orosz}, J.~A., \& {Kuulkers}, E. 1999, \mnras, 305, 132

\bibitem[{{Patruno} \& {Watts}(2012)}]{Patruno2012}
{Patruno}, A., \& {Watts}, A.~L. 2012, ArXiv e-prints

\bibitem[{{Smale}(1998)}]{Smale1998}
{Smale}, A.~P. 1998, \apjl, 498, L141

\bibitem[{{van der Klis}(2006)}]{Van2006}
{van der Klis}, M. 2006, {Rapid X-ray Variability}, ed. W.~H.~G. {Lewin} \&
  M.~{van der Klis}, 39--112

\bibitem[{{Wade} \& {Horne}(1988)}]{Wade1988}
{Wade}, R.~A., \& {Horne}, K. 1988, \apj, 324, 411

\bibitem[{{Watts}(2012)}]{Watts2012}
{Watts}, A.~L. 2012, \araa, 50, 609

\bibitem[{{Watts} {et~al.}(2008){Watts}, {Krishnan}, {Bildsten}, \&
  {Schutz}}]{Watts2008}
{Watts}, A.~L., {Krishnan}, B., {Bildsten}, L., \& {Schutz}, B.~F. 2008,
  \mnras, 389, 839

\bibitem[{{Wijnands} {et~al.}(2003){Wijnands}, {van der Klis}, {Homan},
  {Chakrabarty}, {Markwardt}, \& {Morgan}}]{Wijnands2003}
{Wijnands}, R., {van der Klis}, M., {Homan}, J., {Chakrabarty}, D.,
  {Markwardt}, C.~B., \& {Morgan}, E.~H. 2003, \nat, 424, 44

\bibitem[{{Wijnands} {et~al.}(1998){Wijnands}, {Homan}, {van der Klis},
  {Kuulkers}, {van Paradijs}, {Lewin}, {Lamb}, {Psaltis}, \&
  {Vaughan}}]{Wijnands1998}
{Wijnands}, R., {et~al.} 1998, \apjl, 493, L87

\bibitem[{{Wilson}(1953)}]{Wilson1953}
{Wilson}. 1953, in General Catalogue of Stellar Radial Velocities, 601

\bibitem[{{Yin} {et~al.}(2007){Yin}, {Zhang}, {Zhao}, {Lei}, {Qu}, {Song}, \&
  {Zhang}}]{Yin2007}
{Yin}, H.~X., {Zhang}, C.~M., {Zhao}, Y.~H., {Lei}, Y.~J., {Qu}, J.~L., {Song},
  L.~M., \& {Zhang}, F. 2007, \aap, 471, 381

\end{thebibliography}

\end{document}